# Stimulated Nutation Echo: Dynamic and Decay Properties

G. Bimbo, R. Boscaino,*  M. Cannas, and F.M. Gelardi

*Istituto Nazionale per la Fisica della Materia and Department of Physical and Astronomical Sciences, University of Palermo, Via Archirafi 36, I 90123 Palermo, Italy*

R. N. Shakhmuratov

*Kazan Physical Technical Institute of Russian Academy of Sciences, 10/7 Sibirski trakt, Kazan, 420029 Russia*

ABSTRACT

We study experimentally the dynamical and decay properties of the Stimulated Nutation Echo (SNE) in a two-level spin system. This is the signal which appears in the transient response of the system to the second pulse at time $\tau_1$ elapsed from its beginning and coinciding with the duration of the first pulse. The information about the first pulse duration is imprinted into the population difference of the inhomogeneously broadened ensemble of the two-level absorbers. The decay of the SNE signal has two contributions. One originates from the population decay during the time ô between the pulses. Another is caused by the coherence loss during the excitation by the first pulse and the reading time of the second pulse. Experimental results on the decay properties induced by these mechanisms are presented. The dependence of these decay rates on the pulse intensity is discussed, and its relationship with the anomalous (non-Bloch) decay of other coherent transients in solids is examined.

# I. INTRODUCTION

When a system of two-level centers (atoms or spins) is resonantly driven by the sequence of two pulses, lasting $\tau_1$ and $\tau_2$ and separated by the interpulse distance $\tau$, the response of the system to the second pulse ($t > \tau_1 + \tau$) includes both an initial Rabi-oscillating component and an oscillatory echo, centered at $t = 2\tau_1 + \tau$, namely at a time distance $\tau_1$ from the onset of the second pulse. This echo effect is an example of forced echo, in the sense that it occurs during the refocusing pulse and is emitted by a strongly driven rather than by a freely evolving system. The nutation echo effect originates from the fact that, at the beginning of the second (refocusing) pulse, the system is far from its thermal equilibrium state, as its population difference and its coherences keep memory of the time evolution during the first (preparative) pulse. The non-thermal initial state causes a revival of Rabi oscillations during the second pulse.

Nutation echo was proposed in optical domain in 1979 [1]. Then it was observed in NMR systems [2, 3] and in ESR systems [4-6]. It was studied theoretically and experimentally in great detail in the NMR domain [3], under the name of two-pulse delayed nutation echo. This name was given to emphasize that the delayed nutation used in optical domain for measuring $T_1$-relaxation [7] is the echo phenomenon missed in those experiments because of the poor resolution of the optical detector. It was mentioned as driven echo in a recent review paper [8]. In 1997 the most advanced theory of the echo along with experiments in the optical domain was presented in [9-10] by A. Szabo et al. It is worth noting that in the echo effect considered here the phase coherence and shift between the two exciting pulses is not relevant; in this respect, it is different from the spin and optical rotary echo [11], from nutation time-reversal experiments [12], and from coherent echoes excited by field gradients sequences in NMR imaging [13].

In this paper the nutation echo is investigated in electron spin resonance systems, at microwave frequency, and in such conditions that it originates only from the spectral pattern of the population difference stored in the inhomogeneous spectrum of the system at the end of the first pulse. Following ref. [9,10] we will refer to it as Stimulated Nutation Echo (SNE), in analogy to the three-pulse-stimulated echo that is produced from a hole pattern burnt into an inhomogeneous line. At variance to previous observations of SNE in ESR systems, obtained either in spin systems generated by pulsed light irradiation [4], or by indirect optical detection [5], or by applying Zeeman field pulses [6], in our experiments the SNE is directly excited by pulses of resonant radiation. This configuration yields a higher value of the signal-to-noise ratio, even with respect to previously reported detection in NMR [2,3] and optical systems [9,10].

In particular we report experimental results on the decay properties of the SNE signal. The SNE intensity is expected to decrease on increasing the interpulse distance $\tau$, due to thermalization of the population difference pattern. According to conventional Bloch Equations, this decay should proceed at the rate $T_1^{-1}$. On the other hand, the SNE intensity is expected to decrease as well on increasing the duration $\tau_1$ of the first (exciting) pulse, due to the coherence loss mechanisms. The latter is a major interest point, if one considers that the anomalous decay of the coherent transients in solids is still a matter of debate.

In this regard we recall that the main anomalies found experimentally concern the free-induction decay (FID) [14-19], the hole burning [20] and the Transient Nutation regime [21,22]. The anomalous decay of FID (strongly related to the anomalous width of the hole burnt in the inhomogeneous line) was ascribed to the excitation power dependence of the atomic coherence decay rate. In fact, the experimental results suggest that the actual width of the burnt hole is much less than the value expected on the basis of the conventional Bloch Equations $\sim c\sqrt{T_1/T_2}$, where $c$ is the driving field Rabi frequency, $T_1$ is a population decay time and $T_2$ is the coherence decay time. In several theoretical papers [23-34] (the list is not exhaustive) it was assumed that the dephasing interactions, responsible for the coherence loss, are averaged out by the fast Rabi oscillations of the atom between the ground and excited states, causing $T_2$ to tend to $T_1$ and the spectral narrowing of the hole.

However, the direct measurement of coherence loss during the resonant excitation, carried out using the Transient Nutation regime, revealed the opposite effect, i.e. that the decay rate becomes higher on increasing the driving field amplitude [21,22]. Recent theories [32-34], based on the model of the field-driven relaxation, have been proposed to describe both decay anomalies in the saturation and in the coherent regime.

The paper is organized as follows. In Sect.II we briefly outline the theory of the SNE by Shakhmuratov [9], specializing it to the specific system considered here, to calculate the intensity and the shape of the SNE and its decay properties, hypothesizing the phenomenological Bloch times $T_1$ and $T_2$. In Sect. III we report the experimental results obtained in our ESR system and we compare it with the solutions of the Bloch Equations and with the anomalous decay of the Transient Nutations.

## II. THEORY

In this section we give a brief and simplified sketch of the theory of the SNE presented in ref. [9]. The ensemble of the two-level particles with inhomogeneously broadened absorption line is excited at the line center by the sequence of two pulses of resonant radiation at frequency $\Omega$, of the duration $\tau_1$ and $\tau_2$, separated by $\tau$. The evolution of the resonant and non resonant spins is governed by the Bloch Equations:

$$\dot{u} + \Delta v + u/T_2 = 0 \tag{1}$$

$$\dot{v} - \Delta u - \chi w + v/T_2 = 0 \tag{2}$$

$$\dot{w} + \chi v + (w - w_0)/T_1 = 0 \tag{3}$$

where the Bloch-vector components $u + iv = 2\rho_{12}\exp(-i\Omega t)$ and $w = \rho_{22} - \rho_{11}$ are the combinations of the elements of the particle density matrix $\rho$, $w_0$ is the equilibrium population difference. $T_1$ and $T_2$ are the decay times of the population difference and of the coherence, with $T_1 \gg T_2$. The Rabi frequency $\chi$ is defined by the transition matrix element and the amplitude of the driving field. The parameter $\Delta = \omega - \Omega$ is the detuning of the generic packet (centered at $\omega$) from the resonance and from the line center.

The transient solutions of Eqs.(1-3) were derived by Torrey [35]. They simplify if we assume the strong field limit ($\chi T_2 \gg 1$) and that the first pulse duration $\tau_1$ is much shorter than $T_1$, so that only $T_2$ affects the time evolution of the system during the pulse. Both conditions are well satisfied in our experimental situation. After the preparation pulse, the population difference of the particles is left perturbed:

$$\frac{w(\Delta, \tau_1)}{w_0} = \left(\frac{\chi}{\beta}\right)^2 \left(\frac{\Delta^2}{\chi^2}\exp(-\alpha_1 \frac{\tau_1}{T_2}) + \exp(-\alpha_2 \frac{\tau_1}{2T_2})\cos\beta\tau_1\right) \tag{4}$$

where $\beta = \sqrt{\Delta^2 + \chi^2}$, $\alpha_1 = (\chi/\beta)^2$, $\alpha_2 = 2 - \alpha_1$. The spectral pattern given in Eq.(4) is the ultimate origin of the echo effect we are concerned with. In Figure 1, we report a typical spectral pattern of the population difference, as obtained from Eq.(4), and hypothesizing a Gaussian inhomogeneous lineshape with standard deviation $\sigma$. As shown, at the end of the first pulse, $w(\Delta)$ is strongly modulated inside a region of width $\chi$ around the excitation point (the line center in our case). Obviously, the particular pattern depends both on $\chi$ and on $\tau_1$: on increasing $\chi$ this region broadens and on increasing $\tau_1$ the peaks become closer and closer.

After the first pulse $(t > \tau_1)$, the spins evolve freely for the interpulse distance $\tau$. We assume that $\tau$ is long enough for the complete decay of the coherences $u(\Delta, \tau_1)$ and $v(\Delta, \tau_1)$ induced by the first pulse. Actually, this condition is less restrictive than $\tau \gg T_2$; in fact, as shown in ref. [9], their contribution to the SNE signal decays as $\propto \exp(-\chi\tau)$. So, provided that the interpulse distance $\tau$ much longer than the Rabi period $T_0 = 2\pi/c$, at $t = \tau + \tau_1$, namely at the onset of the second pulse, the memory of the nutational regime excited by the first pulse is stored only in the spectral distribution of the population difference $w(\Delta, \tau_1)$, damped by the longitudinal relaxation:

$$w(\Delta, \tau_1 + \tau) = [w(\Delta, \tau_1) - w_0]\exp(-\tau/T_1) + w_0 \tag{5}$$

$$u(\Delta, \tau_1 + \tau) = 0; \quad v(\Delta, \tau_1 + \tau) = 0$$

These represent the initial conditions of the system at the onset of the second pulse. During the second pulse, the $v$-component of the induced coherence is

$$v(\Delta, t) = \left(\frac{c}{\beta} \sin \beta t\right) \exp\left(-\alpha_2 \frac{t}{T_2}\right) w(\Delta, \tau_1 + \tau) \tag{6}$$

as obtained by solving again Eqs.(1) with the initial conditions in Eq.(5). Here the time $t$ is counted from the beginning of the second pulse. By substituting Eqs (4) and (5), we get:

$$\frac{v(\Delta,t)}{w_0} = \left(\frac{\chi}{\beta}\sin\beta t\right)\exp\left(-\alpha_2\frac{t}{T_2}\right)\left\{\left[1-\exp\left(-\frac{\tau}{T_1}\right)\right] + \frac{\Delta^2}{\beta^2}\exp\left[-\left(\alpha_1\frac{\tau_1}{T_2}+\frac{\tau}{T_1}\right)\right]\right\} + $$
$$+\frac{1}{2}\left(\frac{\chi}{\beta}\right)^3 \exp\left[-\left(\alpha_2\frac{\tau_1+t}{2T_2}+\frac{\tau}{T_1}\right)\right]\left[\sin(\beta(t-\tau_1)) + \sin(\beta(t+\tau_1))\right] \tag{7}$$

For the central excitation of the inhomogeneous absorption line we can ignore the $u(\Delta, t)$ component, since the only nonzero averaged coherence is

$$\langle v(t) \rangle = V(t) = \int_{-\infty}^{+\infty} v(\Delta, t) f(\Delta) d\Delta \tag{8}$$

where $f(\Delta)$ is the distribution function of the resonant frequencies which coincides with the inhomogeneous absorption line in shape. Assuming the line width of the latter is much larger than $\chi$ we can consider the integration in Eq.(8) as the integration with the flat function $f(\Delta) = f(0)$.

In Eq. (7), the amplitude of the term containing $\sin(\beta(t-\tau_1))$ represents the SNE signal $V_{echo}(t)$, centered at $t = \tau_1$. Hereafter, to simplify the calculation we admit $\alpha_1 = \alpha_2 = 1$. Then

$$\frac{V_{echo}(t)}{w_0} = \frac{1}{2} f(0) \exp\left(-\frac{t_1+t}{2T_2} - \frac{t}{T_1}\right) \int_{-\infty}^{+\infty} \left(\frac{c}{b}\right)^3 \sin(b(t-t_1)) \, d\Delta \qquad (9)$$

A typical result of the SNE signal, as calculated using Eq. (9), is reported in the inset of Fig.1. The signal consists of an oscillatory pattern with many spikes of alternating sign. The peak values are distributed inside a symmetric bell-shaped envelope centered at $t = t_1$. The analytical study carried out in Ref.[9] shows that the spikes occur at $t = t_1 \pm t_i$, where the sequence $t_i$ is given in units of the Rabi period $T_0$ as $t_i/T_0 = 0.17; 0.65; 1.24; 1.74; 2.24$, etc. The amplitudes of the spikes decrease as $t_i$ increases. $V_{echo}$ is exactly zero at $t = t_1$.

The SNE becomes pronounced if the first pulse area $q = ct_1$ is at least several ð. Otherwise it is impossible to isolate the echo signal from the transient nutations excited by the second pulse, given by the first term in Eq.(7).

In particular, when varying the relevant time parameters of the exciting sequence, $t_1$ and $t$, the amplitudes $V_m$ of the peaks nearest to the center ($t \approx t_1$) are expected to be damped as:

$$V_m \propto \exp\left(-\frac{t_1}{T_2} - \frac{t}{T_1}\right) \qquad (10)$$

Eq.(10) describes the decay dynamics of the SNE, as expected from the conventional Bloch Equations. In particular, the SNE is expected to decay on increasing the interpulse distance $t$ at a rate fixed by the spin relaxation time $T_1$. However, this decay can be much faster, due to the spectral diffusion, which tends to smear out the distribution of the population difference depicted in Fig.1.

On the other hand, the SNE echo is expected to decay as well as a function of $t_1$ as a single exponential at the rate $1/T_2$, independent of $c$. In this regard we remark that this expectation is not conditioned by the approximation used for $a_2$. In fact, by numerically integrating Eq.(7) using the $D$-dependent expression of $a_2$, we verified that the only effect of the approximation $\alpha_1 = \alpha_2 = 1$ is to overestimate the decay rate (by nearly 10%), in the sense that the actual rate is $1.13/T_2$ rather than $1/T_2$. What is important for our present concern, is that the single-exponential nature of the decay and the $c$-independence of the rate are properties of the exact solution of the Bloch Equations. This is the relevant reason of interest in this coherent regime. In fact, the decay of the SNE as a function of $t_1$ should manifest the processes of coherence loss in the strongly driven system. As mentioned in the previous

Section, in these regimes the decay properties deviate strongly from that predicted by the Bloch Equations. This has been already observed in connection with the Transient Nutation regime and a similar effect is expected to occur for the SNE.

## III. EXPERIMENTS AND DISCUSSION

For the experimental investigation of the SNE in magnetic resonance systems we used the method of the two-photon (TP)-excitation-Second-Harmonic (SH)-detection, already successfully employed for several transient regimes [15, 17, 21]. In our experimental apparatus, the spin system, tuned to the frequency $\omega_o$ ($\omega_o/2\pi$ = 5.95 GHz) by the external static magnetic field $B_o$, is excited by an intense microwave radiation at frequency $\omega = \omega_o/2$, which couples the spin states by TP transitions. The TP-induced coherences manifest in the build-up of a transverse magnetization, $M_\perp(2\omega)$, oscillating at the SH of the driving field. The response of the spin system is monitored by revealing the radiation emitted by the spin system at frequency $2\omega$. Under pulse excitation conditions, the amplitude $M_\perp$ of $M_\perp(2\omega)$ varies in time, reproducing the coherent transient regimes of the spin system. At the exact TP- tuning, $\omega = \omega_o/2$, $M_\perp$ is proportional to V(t) and is the same as for the usual single-quantum excitation, if the appropriate TP-induced Rabi frequency $\chi$ is used [36]. We refer to our previous papers [15, 17, 21] for the detailed description of this procedure and of the related experimental apparatus.

The experiments described below were carried out, at T= 4.2 K, using a sample of glassy $SiO_2$, containing a concentration c ~ 2.4 $10^{17}$ spin/cm$^3$ of E' centers [37], preliminarily generated by exposure to gamma rays [38]. E' centers have S=1/2 and are particularly suitable for this kind of experiments for their relatively long relaxation times. In our work conditions, we measured $T_1$ = 1.2 s (by the saturation recovery method) and $T_2$ = 75 μs (by the spin echo method). In the glassy matrix of $SiO_2$, E' centers exhibit a powder-like line shape. For the purposes of the experiments described below, the central part of their resonance line is well approximated by a Gaussian shape with $\sigma/2\pi$~1 MHz.

The spin system, tuned at resonance, is excited by a sequence of two pulses, lasting $t_1$ and $t_2$ respectively, separated by a distance $\tau$, both with the same power level. The repetition frequency is kept low enough (typically 0.5 Hz) to ensure complete thermal relaxation of the spin system between successive sequences. The microwave signal emitted by the spin system and output by the cavity has a time-dependent amplitude $M_\perp$ (t), which we detect by a phase insensitive superheterodyne receiver to get $|M_\perp(t)|$. To improve the signal-to-noise ratio, $|M_\perp(t)|$ is averaged over typically 16 sequences.

During the first pulse, the system undergoes damped Rabi oscillations, which we use for the accurate measure of $\chi$. A typical response to the second pulse is shown in Fig.2, as detected for $\chi/2\pi = 200$ kHz, $t_1 = 30$ μs ($t_1 \approx 6\ T_o$) and $\tau = 400$ μs ($\tau \gg T_2$). At the onset of the second pulse, the system undergoes fast-damping Rabi oscillations, followed by the SNE regime. The SNE signal consists of a bell-enveloped series of peaks at the Rabi frequency $\chi$, centred at $t_e = 30$μs ($t_e = t_1$) after the onset of the second pulse, in agreement with theoretical predictions.

In order to explore a wide range of $\chi$, $t_1$ and $\tau$, where the SNE signal is nearly at the noise level, we often used a sort of homodyne detection. The microwave signal output by the cavity is firstly superimposed to an equal-frequency reference signal, whose amplitude and phase are adjusted to ensure the maximum visibility of SNE oscillations. The sum signal is then revealed by a conventional superheterodyne detector, yielding $S^2(t) = M_\perp^2 + R^2 + 2 M_\perp R$, where R is the amplitude of the reference signal. The result is shown in the inset of Fig. 2. Note that, usually $R^2 \gg M_\perp^2$, so that $S^2 = R^2 + 2M_\perp R$ consists of a d.c. offset and a time dependent part following the time evolution of $M_\perp(t)$, coherently amplified by R. To estimate the SNE intensity $I_E$, we determine the maximum and the minimum signals nearest to the echo center: $S^2_{max}$ ($= M_\perp^2 + R^2 + 2 |M_\perp| R$) and $S^2_{min}$ ($M_\perp^2 + R^2 - 2 |M_\perp| R$) to get $|M_\perp|^2$ as ($S^2_{max} - S^2_{min}$)/4R.

The detailed timing of the SNE signal is shown in Fig. 3 where we report an expanded view of a typical curve of the SNE signal, taken with a Rabi frequency $\chi = 100$ kHz and $\delta_1 = 55$ μs. In the figure the time origin is placed at 55 μs after the onset of the second pulse and we observe maxima and minima symmetrically located around the origin at times $t_i$, that in units of the Rabi period $T_0$ are: $t_i/T_0 = 0.25$; 0.72; 1.23; 1.74; and 2.27 ($\pm 0.02$, typically), in fair agreement with the theoretical values: 0.17; 0.65; 1.24; 1.74; 2.24.

The echo intensity $I_E$ is expected to decrease on increasing the interpulse distance $\tau$, because of the progressive smearing of the population pattern stored in the longitudinal component of the magnetization. According to the conventional Bloch Equations the decay is expected to proceed at the rate $1/T_1$ [Eq.(10)]. The experimental results are shown in Fig. 4, where the intensity $I_E$ is reported as a function of $\tau$, for various values of $\chi$. As shown, the decay is not a single exponential and occurs in a time scale of a few ms, to be compared to the much longer value of $T_1$ (~ 1.0 s). Moreover, the decay depends on χ, being faster and faster on decreasing χ; on the other hand, apart from the obvious scale factor, the decay curves are essentially independent on the first pulse duration $\boldsymbol{t}_1$.

The spectral diffusion (SD) may be the origin of so fast a decay. In fact the migration of the excitations over the whole resonance line may affect the persistence of the population spectral pattern stored during the first pulse. The SD process is usually faster than spin-lattice relaxation and may be the dominant cause of echo decay. In this regard we note that the excitation by the first pulse is spectrally selective in two aspects. On the one hand it involves only a narrow part of the resonance line (with a width of the order of χ << σ). This situation is usually described in terms of A-type (echo-active) spins and B-type (the others) spins [39]. The dipolar interaction A-B drags spin excitations away from the line centre and causes the decay of the echo. On increasing χ, the number of A-type spins increases with respect to B-type spins; this circumstance may account for the slower decay of the echo at high χ, experimentally observed. On the other hand, the first pulse excitation is spectrally selective also within the A-type spin group. In fact it establishes a comb population pattern (Fig.1) within the central part of the line, which is the true source of the SNE formation. So, one expects that the A-A dipolar interaction may smear this pattern and yield an additional contribution to the echo decay. The population pattern is finer and finer on increasing $\boldsymbol{t}_1$, with higher spectral gradients of the population difference. So one expects this contribution to the decay rate to increase on increasing $\boldsymbol{t}_1$. The experimental results suggest that the former effect (due to A-B interaction) is predominant.

The echo intensity $I_E$ is expected to decay as well on increasing the first pulse duration $\boldsymbol{t}_1$, while keeping fixed χ and τ, due to the coherence loss occurring during both pulses, for a total time span $2\boldsymbol{t}_1$. The experimental results are reported in Fig.5, where we report the echo intensity $I_E$, measured as described above, as a function of $\boldsymbol{t}_1$, for three different values of the Rabi frequency χ, 50 kHz, 100 kHz and 200 kHz. These results are unaffected by the interpulse distance τ. In the same figure we report also the decay function, as predicted by the solutions of the conventional Bloch Equations: according to Eq.(10), the decay of the echo intensity near the centre (t ≈ $\boldsymbol{t}_1$) should follow a single exponential law with the rate $T_2^{-1}$. We note that the experimental decay is well described by a single exponential but proceeds at a rate faster than $T_2^{-1}$ and power-dependent. The experimental dependence of $\Gamma_E$ on $\boldsymbol{c}$ is reported in Fig. 6. As shown, it is well described by a linear law and is compatible with the Bloch value for $\chi$ tending to zero. By fitting we get $\Gamma_E = \alpha + \beta\chi$, with $\alpha/2\pi = (1.0 \pm 0.2)$ kHz and $\beta = (1.3 \pm 0.2)\, 10^{-2}$.

The $\chi$-dependence of the experimental rate $\Gamma_E$ of the SNE decay is reminiscent of the non-Bloch decay of the Transient Nutation regime, already observed in the same system for single pulse excitation. In those experiments it was found that the decay rate $\Gamma_{TN}$ of TN tends to $(2T_2)^{-1}$ in the limit of very low power but increases linearly with $\chi$. The experiments reported here show that the decay rate $\Gamma_E$ of the SNE follows a qualitatively similar behaviour, confirming that the coherence loss effect of strongly driven systems is not adequately described by the Bloch model.

In the inset of the figure, for the sake of comparison, we report the experimental $\chi$-dependence of the $\Gamma_{TN}$ as measured in the same sample and well fitted by the linear law $\Gamma_{TN} = \alpha + \beta'\chi$ with $\beta' = (10.6 \pm 0.5)\, 10^{-2}$. So, the two behaviours are similar, but the slopes are different, in the sense that the $\chi$-sensitivity of $\Gamma_E$ seems to be less than $\Gamma_{TN}$. A qualitative reason for this difference may be found recalling the origin of the anomalous decay of the coherent transient regime. As discussed in [32, 34], the anomalous behaviour of the coherence loss rate may be ascribed to the modifications of the dipolar field caused by the resonant interaction with the microwave field, which yields a linear dependence of the rate $\Gamma_{TN}$ on *c*. Obviously, the same mechanism is expected to be effective also during the second refocusing pulse. However, the effect may be less due to the population spectral pattern which reduces the number of the spins involved in the echo formation with respect to those involved in the Transient Nutation regime, thus lessening the part of the dipolar interaction modified by the resonant field.

## IV. CONCLUSION

In conclusion, we have reported experimental results on the SNE in electron spin resonance systems. The timing and the shape of the echo pattern appear in agreement with the theory by Szabo and Shakhmuratov [9]. We have investigated the decay properties of the echo and found that the decay is strongly affected by the spectral diffusion mechanism; our results are consistent with the prevalence of the A-B interaction over the A-A. Moreover, we studied the effect of the coherence loss during the exciting and the refocusing pulses. Our results evidence a non-Bloch behaviour similar to that observed in other coherent regimes which we tentatively ascribe to the field-induced modification of the dipolar field. This may be the origin as well of the different power dependence of coherence loss rate measured in the simple Transient Nutation regime and in the SNE regime. In this regard we note that the experimental study of the SNE echo excited by two pulses with different $\chi$ may yield further information. Work is in progress in this direction.

## ACKNOWLEDGMENTS

Partial financial support has been given by the Italian Ministry for University and Scientific Research, by University of Palermo, Italy (International Cooperation Agreement Palermo-Kazan). Technical assistance by G. Lapis is gratefully acknowledged. RNS acknowledges support from INTAS and Russian Fund for Basic Research.

**REFERENCES**

1. A. I. Alekseev and A. M. Basharov, Zh. Eksp. Teor. Fiz. **77**, 537 (1979) [Sov. Phys. JETP **50**, 272(1979)].

2. C. Bock, M. Mehring, H. Seidel, and H. Weber, Proc. of the Joint ISMAR-Ampere International Conference on Magnetic Resonance, Delft, 1980 (Ed. J. Smidt and W.H.Wisman, The Franklin Institute Press) p. 421(1980).

3. V. S. Kuz'min, A. P. Saiko, and G. G. Fedoruk, Zh. Eksp. Teor. Fiz. **99**, 215 (1991) [Sov. Phys. JETP **72**, 121 (1991)].

4. R. Furrer, F. Fujara, C. Lange, D. Stehlik, H. M. Vieth, and W. Vollmann, Chem. Phys. Lett. **75**, 332 (1980).

5. A. M. Ponte Goncalves and R. Gillies, Chem. Phys. Lett. **94**, 21 (1983).

6. G.G. Fedoruk, J. Appl. Spectroscopy **65**, 420 (1998)

7. J. Schmidt, P. R. Berman and R. G. Brewer, Phys. Rev. Lett. **31**, 1103 (1973); P. R. Berman, J. M. Levy and R. G. Brewer, Phys. Rev. A **11**, 1668 (1975); R. G. Brewer and A. Z. Genack, Phys. Rev. Lett. **36**, 959 (1976); A. Schenzle and R. G. Brewer, Phys. Rev. A **14**, 1756 (1976); R. L. Shoemaker, in *Laser and Coherence Spectroscopy* (Ed. by J. I. Steinfeld), Plenum Press, New York (1978), p. 197.

8. A. Ponti and A. Schweiger, Appl. Magn. Reson. **7**, 363 (1994).

9. A. Szabo and R. N. Shakhmuratov, Phys. Rev. A **55**, 1423 (1997).

10. A. Szabo and R. N. Shakhmuratov, Bulletin of the Russian Academy of Sciences, Physics **62**, 217 (1998) [Izvestiya Rossiiskoi Academii Nauk, Seriya Fizicheskaya **62**, 261 (1998)].

11. I. Solomon, Phys. Rev. Lett. **2**, 301 (1959); N.C. Wong, S.S. Kano, R.G. Brewer, Phys. Rev. A **21**, 260 (1980).

12. Y.S.Bai, A.G.Yodh, and T.W. Mossberg, Phys. Rev. A **34** 1222 (1986).

13. I. Ardelean, R. Kimmich, A. Klemm, J. Mag. Res. **146** 43 (2000)

14. R.G. DeVoe and R.G. Brewer, Phys. Rev. Lett. **50**, 1269 (1983).

15. R. Boscaino, F.M. Gelardi, and G. Messina, Phys. Rev. A **28**, 495 (1983).

16. A. Szabo and T. Muramoto, Phys. Rev. A **39**, 3992 (1989).

17. R. Boscaino and V.M. La Bella, Phys. Rev. A **41**, 5171 (1990).

18. R. Boscaino and F.M. Gelardi, Phys. Rev. A **45**, 546 (1992).

19. R.N. Shakhmuratov and A. Szabo, Phys. Rev. B **48**, 6903 (1993).

20. A. Szabo and R. Kaarli, Phys. Rev. B **44**, 12307 (1991).

21. R. Boscaino, F.M. Gelardi, and J.P.Korb, Phys. Rev. B **48**, 7077 (1993).

22. S. Agnello, R. Boscaino, M. Cannas, F.M. Gelardi, and R.N. Shakhmuratov, Phys. Rev. A **59**, 4087 (1999).

23. E. Hanamura, J. Phys. Soc. Jpn. **52**, 2258 (1983); *ibid* **52**, 3678 (1983).

24. A. Schenzle, M. Mitsunaga, R.G. DeVoe, and R.G. Brewer, Phys. Rev. A **30**, 325 (1984).

25. J. Javanainen, Opt. Commun. **50**, 26 (1984).

26. P.A. Apanasevich, S.Ya. Kilin, A.P. Nizovtsev, and N.S. Onishchenco, Opt. Commun. **52**, 279 (1984); J.Opt.Soc.Am. B **3**, 587 (1986).

27. J.H. Eberly, K. Wodkiewicz, and B.W. Shore, Phys. Rev. A **30**, 2381 (1984); K. Wodkiewicz and J.H. Eberly, Phys. Rev. A **31**, 2314 (1985); *ibid* **32**, 992 (1985).

28. P.R. Berman and R.G. Brewer, Phys. Rev. A **32**, 2784 (1985); P.R. Berman, J.Opt.Soc.Am. B **3**, 572 (1986).

29. A.R. Kessel, R.N. Shakhmuratov, and L.D. Eskin, Zh. Eksp. Teor. Fiz. **94**, 202 (1988) [Sov. Phys. JETP **67**, 2071 (1988)]; R.N. Shakhmuratov, Pis'ma Zh. Eksp. Teor. Fiz. **51**, 454 (1990) [JETP Lett. **51**, 513 (1990)].

30. A.I. Burshtein, A.A. Zharikov, and V.S. Malinovsky, Zh. Eksp. Teor. Fiz. **96**, 2061 (1989) [Sov. Phys. JETP **69**, 1164 (1989)]; Phys. Rev. A **43**, 1538 (1991); A.I. Burshtein and V.S. Malinovsky, J.Opt.Soc.Am. B **8**, 1098 (1991); V.S. Malinovskii, Opt.Spektrosk. **70,** 681 (1991) [Opt. Spectrosc (USSR) **70**, 399 (19919)]; Phys.Rev. A **52**, 4921 (1995).

31. R.N. Shakhmuratov and A. Szabo, Laser Phys. **3**, 1042 (1993); Phys. Rev. A **58**, 3099 (1998).

32. R.N. Shakhmuratov, F.M. Gelardi, and M. Cannas, Phys. Rev. Lett. **79**, 2963 (1997); SPIE Proceedings (USA) **3239**, 206 (1997).

33. R. N. Shakhmuratov, Phys. Rev. A 59, 3788 (1999).

34. N. Ya. Asadullina, T. Ya. Asadullin, Ya. Ya. Asadullin, J. Phys.: Condens. Matter **13**, 3475; *ibid* 5231 (2001)

35. H. C. Torrey, Phys. Rev. **76**, 1059 (1949).

36. R. Boscaino and G. Messina, Physica C **138**, 179 (1986)

37. R.A. Weeks and C.M. Nelson, J. Am. Ceram. Soc. **43**, 389 (1960); K.L. Yip and W.B. Fowler, Phys. Rev. B **11**, 2327 (1975); D.L. Griscom, *ibid.* **20**, 1823 (1979).

38. R. Boscaino, M. Cannas, F.M. Gelardi, and M. Leone, Nucl. Instrum. Methods Phys. Res. B **116**, 373 (1996).

39. R. Boscaino, F.M. Gelardi, and M. Cannas, Phys. Rev. B **53**, 302 (1996); S. Agnello, R. Boscaino, M. Cannas, F.M.Gelardi, *ibid.* **64**, 174423 (2001).

**FIGURE 1.** Spectral pattern of the population difference at the end of the first pulse. The dashed line plots the Gaussian inhomogeneous line with standard deviation $\sigma$. The full line plots the population difference, as calculated from Eq.(4), with $\chi/\sigma = 0.2$, $\tau_1/T_0 = 2$, $T_2/T_0 = 2000$. The dotted line is the envelope of the hole burnt within the inhomogeneous line. The inset shows the corresponding SNE signal: time $t$ is measured from the beginning of the second pulse and is given in units of the Rabi period $T_0$.

**FIGURE 2.** Typical SNE signal, as detected after a preparative pulse with $\chi/2\pi = 200$ kHz and lasting $\tau_1 = 30$ μs. Time t is measured from the onset of the second pulse, 400 μs after the first one. The same signal is shown in the inset, as detected using the superposition to a reference signal, as described in the text.

**FIGURE 3.** Timing of the SNE signal, taken with $\chi = 2\pi$ 100 kHz, $\tau_1 = 55$ μs, $\tau = 40$ μs. The time origin is placed at 55 μs after the onset of the second pulse and coincides with the center of the SNE signal. Dashed lines indicate the maxima and minima, occurring at the values $t_i$ listed in the text.

**FIGURE 4.** Decay of the SNE signal on varying the interpulse distance $\tau$, at various values (indicated in the figure) of the Rabi frequency $\chi$. Symbols are experimental values of SNE intensity, measured as described in the text. Lines are only guides for eyes. The curves are shifted vertically to optimise the visibility.

**FIGURE 5.** Decay of the SNE signal on varying the duration $t_1$ of the first (exciting) pulse, for three values of the Rabi frequency, indicated in the figure. The interpulse distance is $t = 150$ μs. The dashed lines are the exponential laws that best the experimental points and are used to measure $\Gamma_E$. The full line represents the decay curve expected on the basis of the Bloch Equation.

**FIGURE 6.** Dependence of the measured decay rate $\Gamma_E$ on the Rabi frequency $\chi$. The full line is the linear best fit to the experimental points. The intercept, $(1.0 \pm 0,2)$ kHz, is consistent with the value $\Gamma_{BE} = (2 T_2)^{-1}$ expected in the Bloch limit. The best fit slope is $\beta = (1.3 \pm 0.2) 10^{-2}$. The inset reports the experimental ÷-dependence of the Transient Nutation decay in the same sample (data taken from ref.[22]).

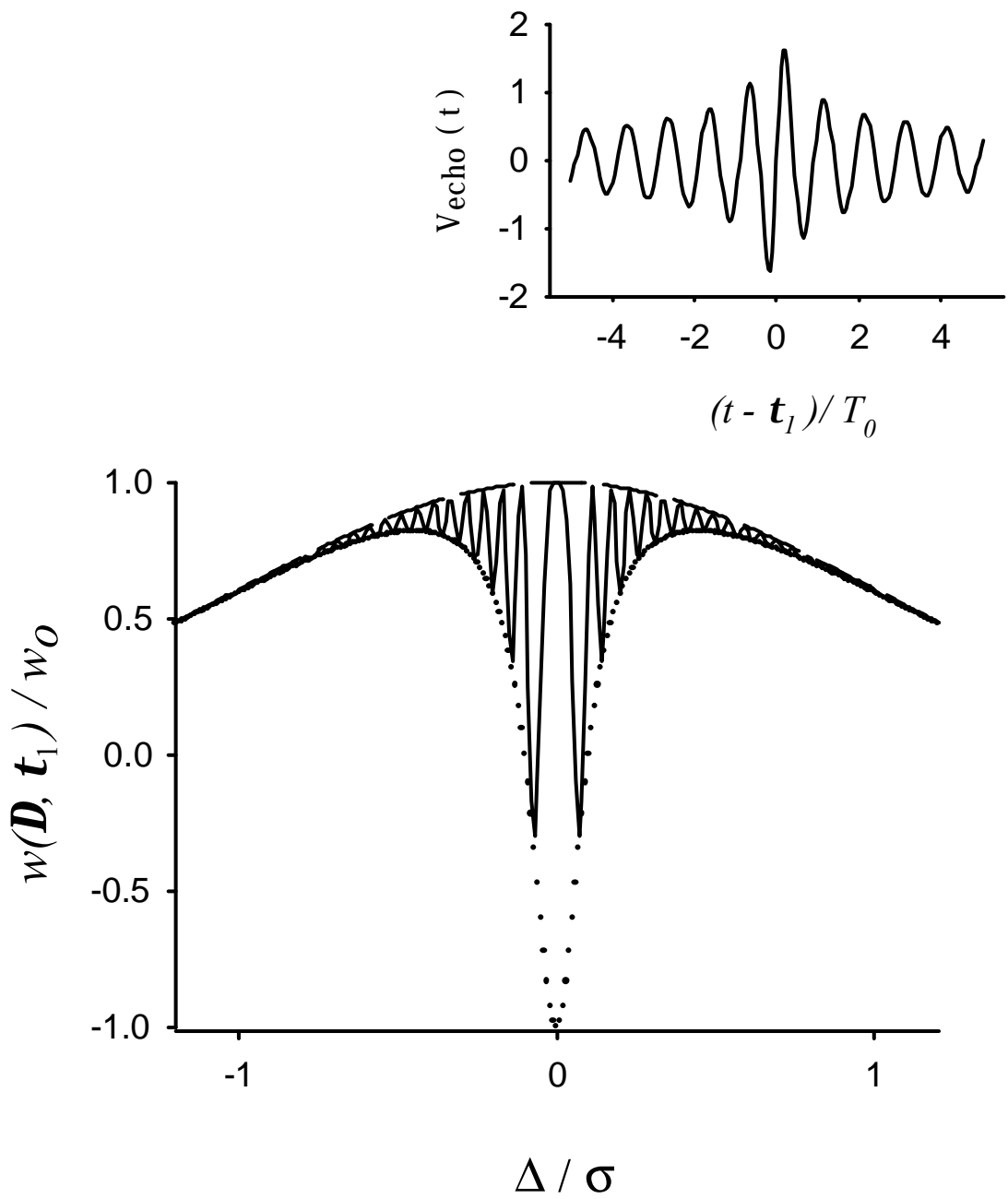

Figure 1
Bimbo et al

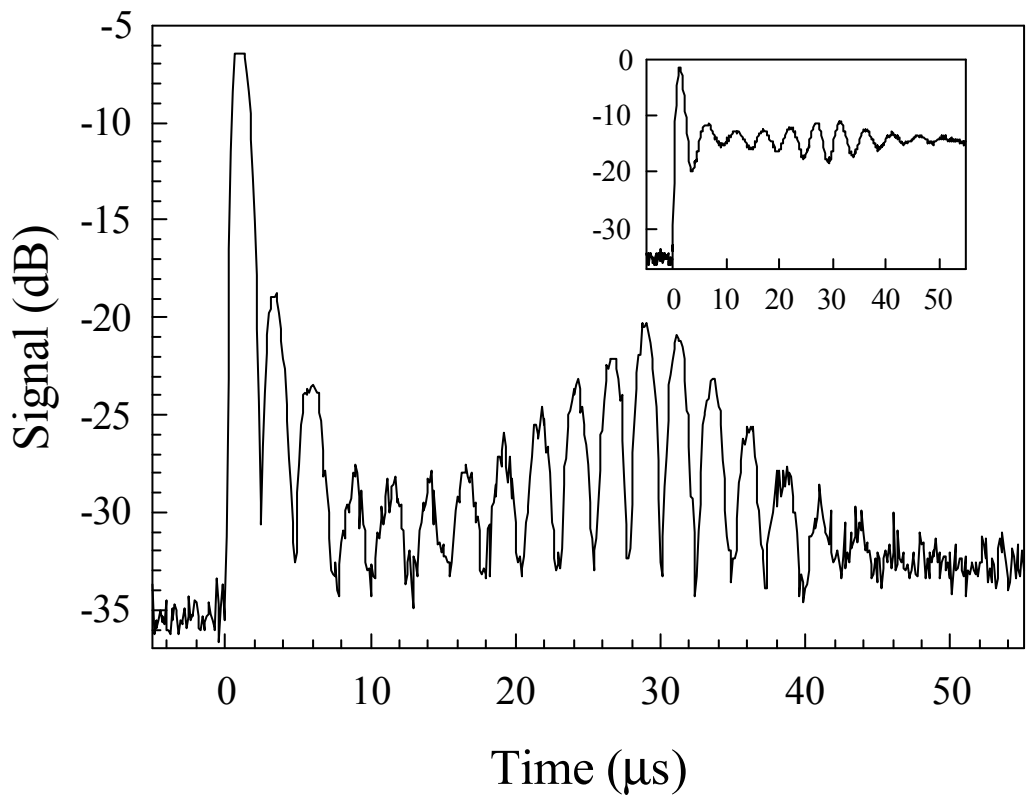

Figure 2
Bimbo et al

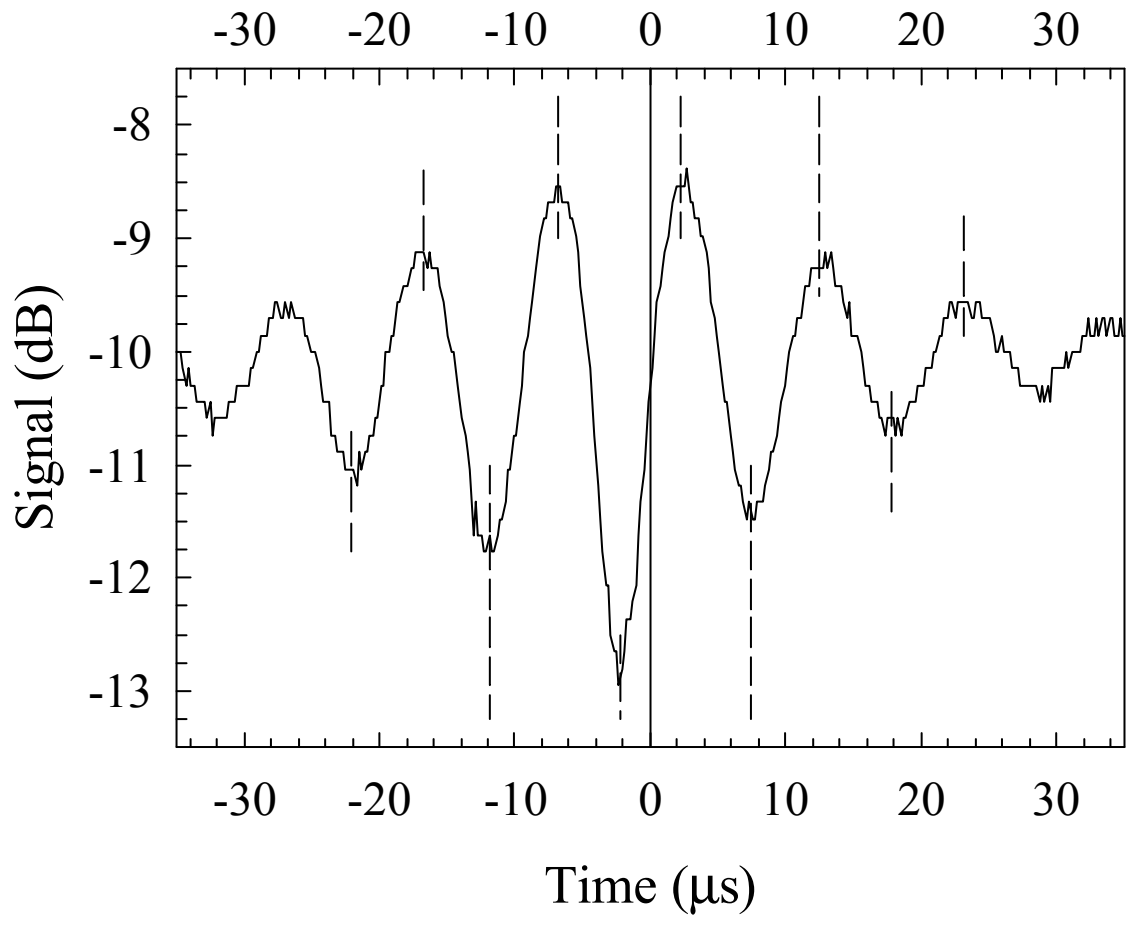

Figure 3
Bimbo et al

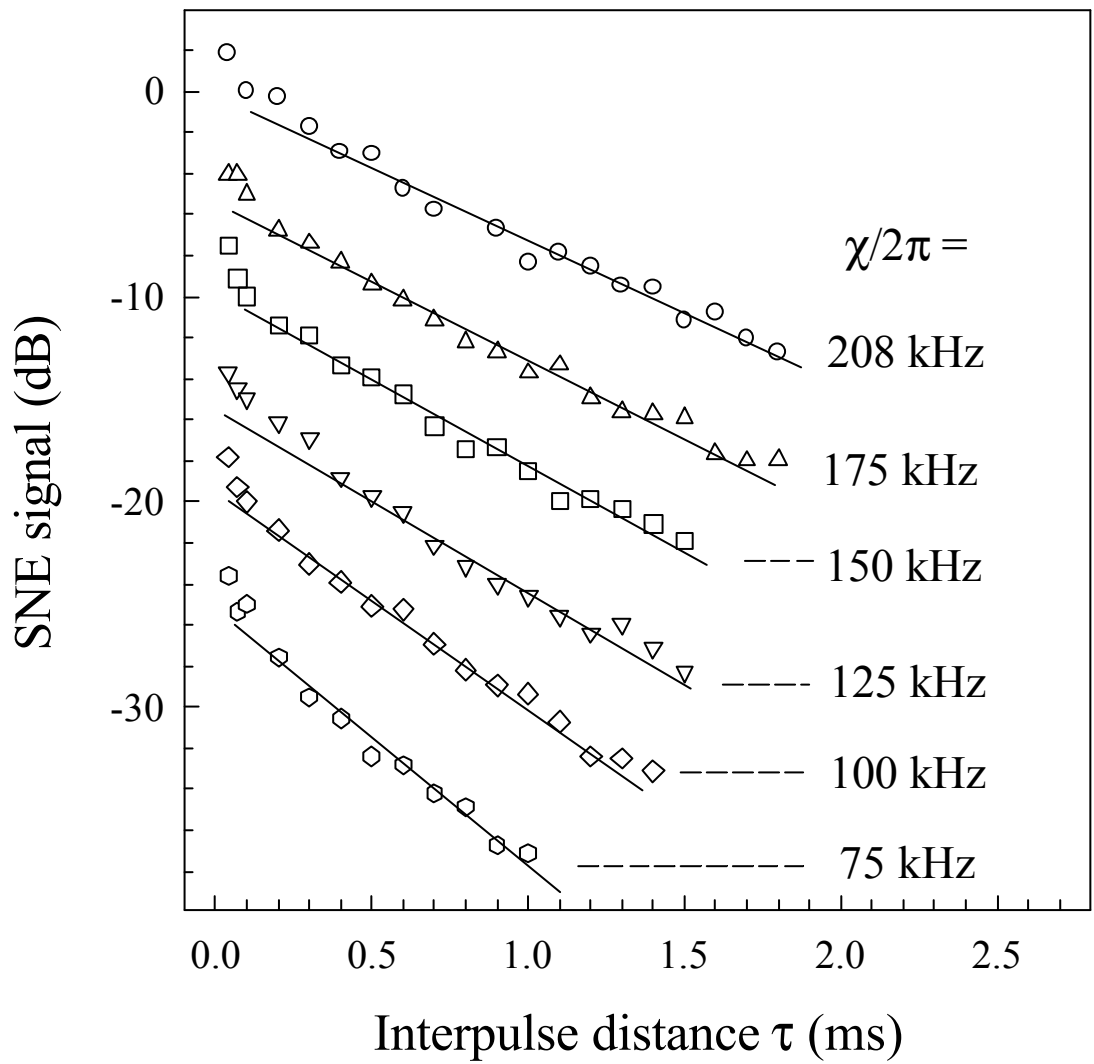

Figure 4
G. Bimbo et al

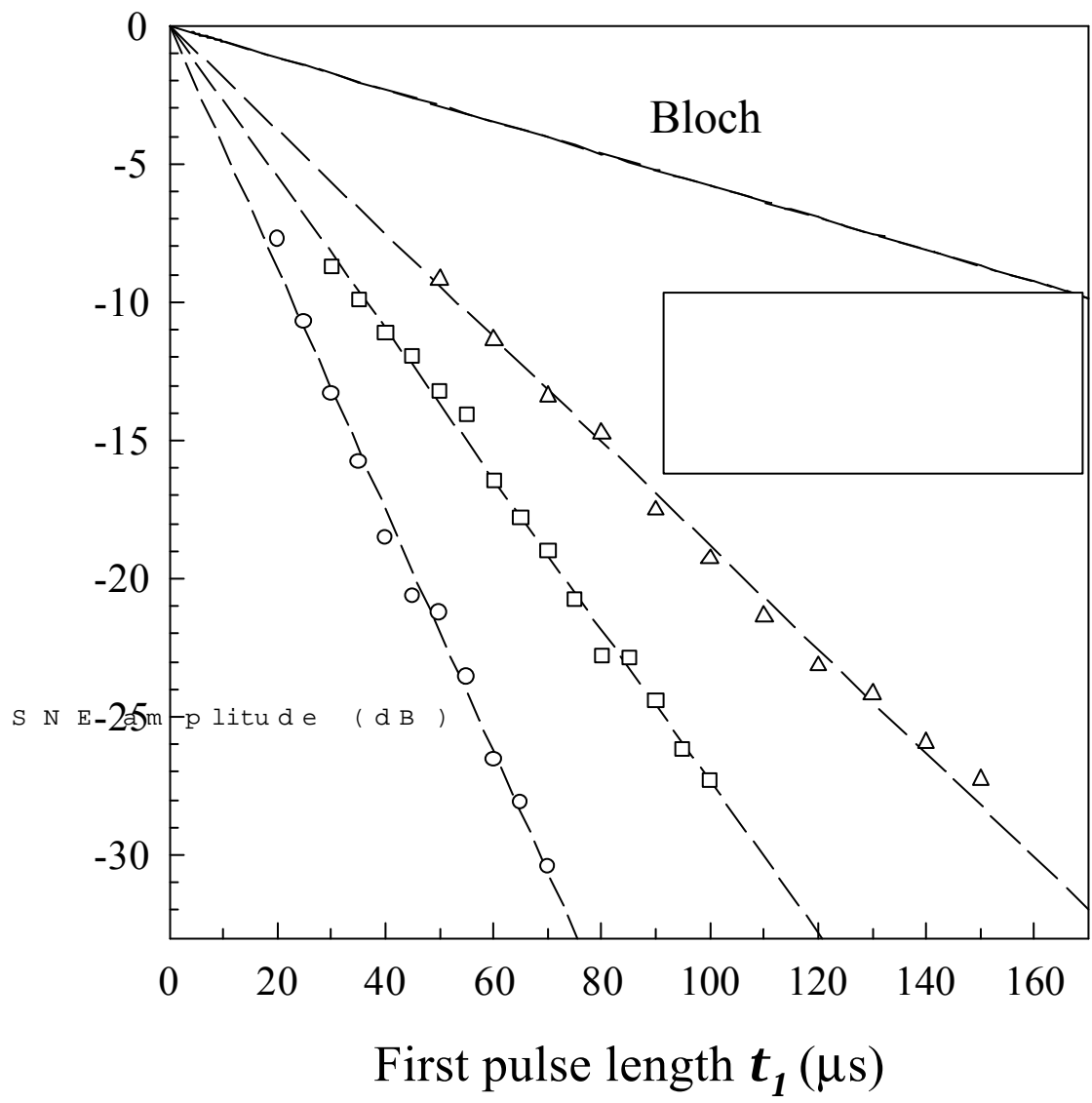

Figure 5
Bimbo et al

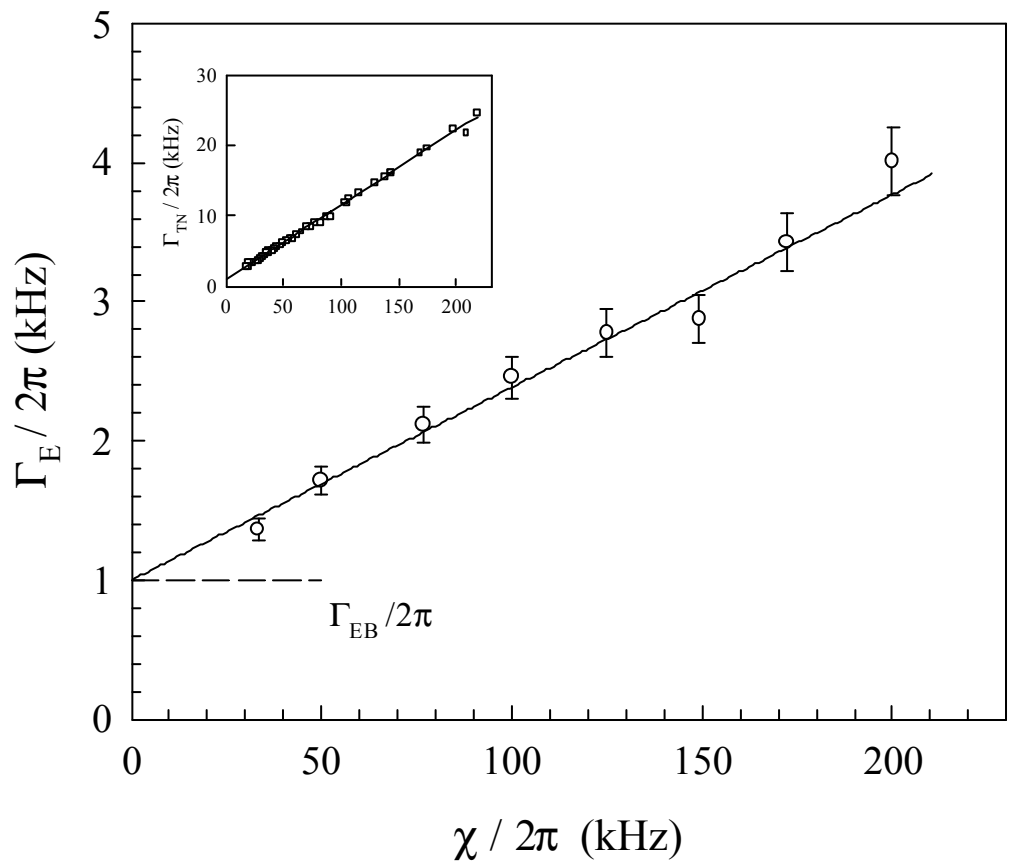

Figure 6
Bimbo et al